\documentclass[traditabstract,a4]{aa}  

\usepackage{epsfig}
\usepackage{graphicx}
\usepackage{epstopdf}
\usepackage{color}  
\usepackage{array}   
\usepackage{units}  
\usepackage[breaklinks, colorlinks, citecolor=blue]{hyperref}
\usepackage{natbib}  
\usepackage{multirow}   
\usepackage{amsmath,amssymb}  
\usepackage[english]{babel}
\usepackage[latin1]{inputenc}
\usepackage[T1]{fontenc}
\usepackage{longtable,lscape}
\usepackage{footnote}
\usepackage{txfonts}
\usepackage[toc,page]{appendix} 

\begin{document}


   \title{High significance detection of the tSZ effect relativistic corrections} 

   \author{G. Hurier\inst{1,2}
          }

\institute{Centro de Estudios de F\'isica del Cosmos de Arag\'on (CEFCA),Plaza de San Juan, 1, planta 2, E-44001, Teruel, Spain \and Institut d'Astrophysique Spatiale, CNRS (UMR8617) Universit\'{e} Paris-Sud 11, B\^{a}timent 121, Orsay, France
\\
\\
\email{ghurier@ias.u-psud.fr} 
}

   \date{Received /Accepted}
 
   \abstract{
   The thermal Sunyaev-Zel'dovich (tSZ) effect is produced by the interaction of cosmic microwave background (CMB) photons with the hot (a few keV) 
   and diffuse gas of electrons inside galaxy clusters integrated along the line of sight.
   This effect produces a distortion of CMB blackbody emission law. This distortion law depends on the electronic temperature of the intra-cluster hot gas, $T_{e}$, through the so-called tSZ relativistic corrections.
   In this work, we have performed a statistical analysis of the tSZ spectral distortion on large galaxy cluster samples. We performed a stacking analysis for several electronic temperature bins, using both spectroscopic measurements of X-ray temperatures and a scaling relation between X-ray luminosities and electronic temperatures.
   We report the first high significance detection of the relativistic tSZ at a significance of 5.3 $\sigma$.  We also demonstrate that the observed tSZ relativistic corrections are consistent with X-ray deduced temperatures.
   This measurement of the tSZ spectral law demonstrates that tSZ effect spectral distorsion can be used as a probe to measure galaxy cluster temperatures.}


   \maketitle


\section{Introduction}

Galaxy clusters contain a hot thermal plasma that Comptonize the cosmic microwave background (CMB) photons when they are traversing the galaxy cluster.
This interaction produces the well known thermal Sunyaev-Zel'dovich (tSZ) effect \citep{sun72}, which induces a spectral distortion of the CMB blackbody emission law.
This spectral distortion can be considered as independent of the electron energies as long as the electron velocities are significantly smaller than the speed of light, $c$. 
Typical plasma inside galaxy clusters have a temperature of several keV. 
Thermal electrons in hot galaxy clusters ($\simeq 5$ keV) have velocities of the order of 0.1 $c$.
Consequently, relativistic corrections have to be applied to the tSZ spectral distorsions \citep{wri79}. 
Fitting formula have been proposed to ease the modeling of the tSZ relativistic corrections \citep{ito00}.\\

These relativistic corrections offer the possibility of using the tSZ effect spectral distorsion as a probe to measure the temperature of the hot plasma inside galaxy clusters \citep{poi98,ens04}.
Recent works \citep{zem10,zem12} have reported evidence of the tSZ relativistic corrections up to 3 $\sigma$ using Z-Spec.\\

Galaxy clusters host galaxies, and consequently the tSZ effect is spatially correlated with radio and infrared emission from galaxies.
Radio galaxies and infrared emissions have been shown to be a significant bias for tSZ-based studies \citep{hur13, planckszcib}.
Thus, if not considered carefully these emissions could significantly bias an attempt of detection for the tSZ relativistic corrections.\\

The $T_{\rm CMB}$ measurement using tSZ effect by \citet{hur14} has shown that Planck fullsky multi-wavelength 
observations of the sub-millimeter and microwave skies are tailored to study the properties of the tSZ spectral distorsions.
In this work, we present the first high-significance detection of the tSZ relativistic correction using a statistical analysis of a large galaxy cluster sample. The paper is organized as follows, Section~\ref{sec_data} presents the data used in this analysis, Section~\ref{sec:theory} describes the tSZ effect, Section~\ref{sec_meth} describes the methodology, and Section~\ref{sec_ana} presents the results.

\section{The data}
\label{sec_data}

\subsection{Planck intensity full-sky maps}
\label{planckmaps}

This paper uses the first 15.5 month survey mission of Planck HFI \citep{PlanckEMI},
corresponding to two full-sky surveys \citep{PlanckMIS}.  We refer to
\citet{PlanckDPC} and \citet{PlanckCAL} for the generic scheme of time-ordered information (TOI) processing and map-making, as well as for
the technical characteristics of the Planck frequency maps.  The
Planck channel maps are provided in {\tt HEALPix} \citep{gor05}
$N_{\mathrm{side}}=2048$ at full resolution. An error map is
associated with each channel map and is obtained from the difference
of the first half and second half of the survey rings, for a given
pointing position of the satellite spin axis.  Here we approximate the Planck HFI beams by effective
circular Gaussians with FWHM up to 5 arc minutes that can be found in \citet{PlanckBEAM}.

\subsection{Catalogs}
\label{plcat}
We used two different galaxy cluster samples. The first one considers all galaxy clusters in the MCXC catalog \citep{pif11}. For this sample we computed the electron temperatures using the scaling relation from \citet{pra09},
\begin{align}
L_X = (0.079 \pm 0.008)\, (T[{\rm keV}])^{2.70 \pm 0.24} \, 10^{44} {\rm erg.s^{-1}}.
\end{align}

In the second sample we considered galaxy clusters for which we have a spectroscopic temperature \citep{cav08,zha08,vik09,pra09,eck11,mit11,rei11,mah13,lag13}.\\
In Table.~\ref{tab_cat}, we summarize the main characteristics of each galaxy cluster catalogs, $N_{\rm cl}$ is the number of objects in the catalog,
  ${T_{e,{\rm min}}}$ and ${T_{e,{\rm max}}}$ are the covered range of temperature, $T_{e,{\rm med}}$ is the median.
  We stress that the different catalogs present overlaps of objects. This overlap has been considered in the following analysis.

\begin{savenotes}
\begin{table}
\caption{Main characteristics of the galaxy cluster catalogs. $N_{\rm cl}$ is the number of objects in the catalog,
  ${T_{e,{\rm min}}}$ and ${T_{e,{\rm max}}}$ are the covered range of temperature, $T_{e,{\rm med}}$ is the median, ${z_{{\rm min}}}$ and ${z_{{\rm max}}}$ are the covered range of  redshift and $z_{\rm med}$ is the median redshift}
\label{tab_cat}
\centering
\begin{tabular}{|c|c|c|c|c|c|c|c|}
\hline
Catalog & $N_{\rm cl}$ & ${T_{e,{\rm min}}}$ & ${T_{e,{\rm max}}}$ & ${T_{e,{\rm med}}}$ & ${z_{{\rm min}}}$ & ${z_{{\rm max}}}$ & ${z_{{\rm med}}}$  \\
\hline
MCXC & 1743 & 0.11 & 10.25 & 2.81 & 0.00 & 1.26 & 0.14 \\
CAV08 & 192 & 2.44 & 19.13  & 6.99 & 0.03 & 1.24 & 0.26\\
ZHA08 & 37 & 3.2 & 11.6 & 6.7 & 0.14 & 0.30 & 0.23 \\
VIK09 & 85 & 2.13 & 14.72 & 4.4 & 0.03 & 0.89 & 0.09 \\
PRA09 & 31 & 2.07 & 8.91 & 3.85 & 0.06 & 0.18 & 0.12 \\
ECK11 & 26 & 0.62 & 2.99 & 1.61 & 0.01 & 0.05 & 0.02 \\
MIT11 & 64 & 0.90 & 15.91 & 4.37 & 0.00 & 0.22 & 0.05 \\
REI11 & 232 & 2.0 & 15.0 & 5.9 & 0.04 & 1.46 & 0.30 \\
MAH13 & 50 & 3.1 & 12.1 & 6.5 & 0.15 & 0.55 & 0.24 \\
LAG13 & 117 & 2.0 & 15.2 & 5.8 & 0.11 & 1.24 & 0.35 \\
\hline
\end{tabular}
\end{table}
\end{savenotes}

\section{The tSZ effect}
\label{sec:theory}

The thermal Sunyaev-Zel'dovich effect \citep{sun72} is a distortion of the CMB blackbody radiation through inverse Compton scattering. CMB photons receive an average energy boost by collision with hot (a few keV) ionized electrons of the intra-cluster medium \citep[see e.g.,][for reviews]{bir99,car02}.
The  thermal SZ Compton parameter in a given direction, $\vec{n}$, on the sky is given by
\begin{equation}
y (\vec{n}) = \int n_{e} \frac{k_{\rm{B}} T_{\rm{e}}}{m_{\rm{e}} c^{2} } \sigma_{T} \  \rm{d}s
\label{comppar}
,\end{equation}
where d$s$ is the distance along the line-of-sight, $\vec{n}$, and $n_{\rm{e}}$
and $T_{e}$ are the electron number density and temperature,
respectively.
In units of CMB temperature, the contribution of the tSZ effect for a given observation frequency $\nu$ is
\begin{equation}
\frac{\Delta T_{\rm{CMB}}}{T_{\rm{CMB}} }= g(\nu) \ y.
\end{equation}
Neglecting relativistic corrections we have 
\begin{equation}
g(\nu) = \left[ x\coth \left(\frac{x}{2}\right) - 4 \right],
\label{szspec}
\end{equation}
with $ x=h \nu/(k_{\rm{B}} T_{\rm{CMB}})$. At $z=0$, where $T_{\rm CMB}(z=0)$~=~2.726$\pm$0.001~K, the tSZ effect is negative below 217~GHz and positive for higher frequencies.\\
Compton parameter to CMB temperature, K$_{\rm{CMB}}$, conversion factors for each frequency channel depend of the convolution of this tSZ contribution to the sky intensity with the Planck frequency responses.\\

This characteristic spectral signature of tSZ effect makes it a unique tool for the detection of galaxy clusters as presented in \citet{planckpsz} and is related to $T_{\rm e}$ through relativistic corrections.\\
The relativistic corrections on the tSZ emission law have been computed as presented in \citep{poi98}.
From this estimation, if we assume that the relativistic corrections on the tSZ emission law can be described as a first order approximation \citep[see][for a detailed fitting formula]{ito00},
\begin{equation}
\Delta T^{\rm relat}_{\rm CMB}(T_e) = \Delta T^{\rm unrelat}_{\rm CMB} + T_e \Delta T^{\rm cor}_{\rm CMB}, 
\label{linap}
\end{equation}
the averaged tSZ emission from different electron populations at various temperatures can be modeled with a single temperature. This approach enables the possibility to perform stacking analyses of the tSZ relativistic corrections.
This approximation is already implicitly considered when fitting a single temperature on the observed tSZ signal. Considering that electronic temperature varies along the line of sight. We stress that the quasi linear behavior of the tSZ spectral distorsion relativistic corrections with respect to $T_e$ is only used to motivate a stacking analysis. In the following, when fitting for $T_e$ on the stacked tSZ signal we use the exact tSZ spectral distorsion as a function of $T_e$.\\

\begin{figure}[!th]
\center
\includegraphics[width=9cm]{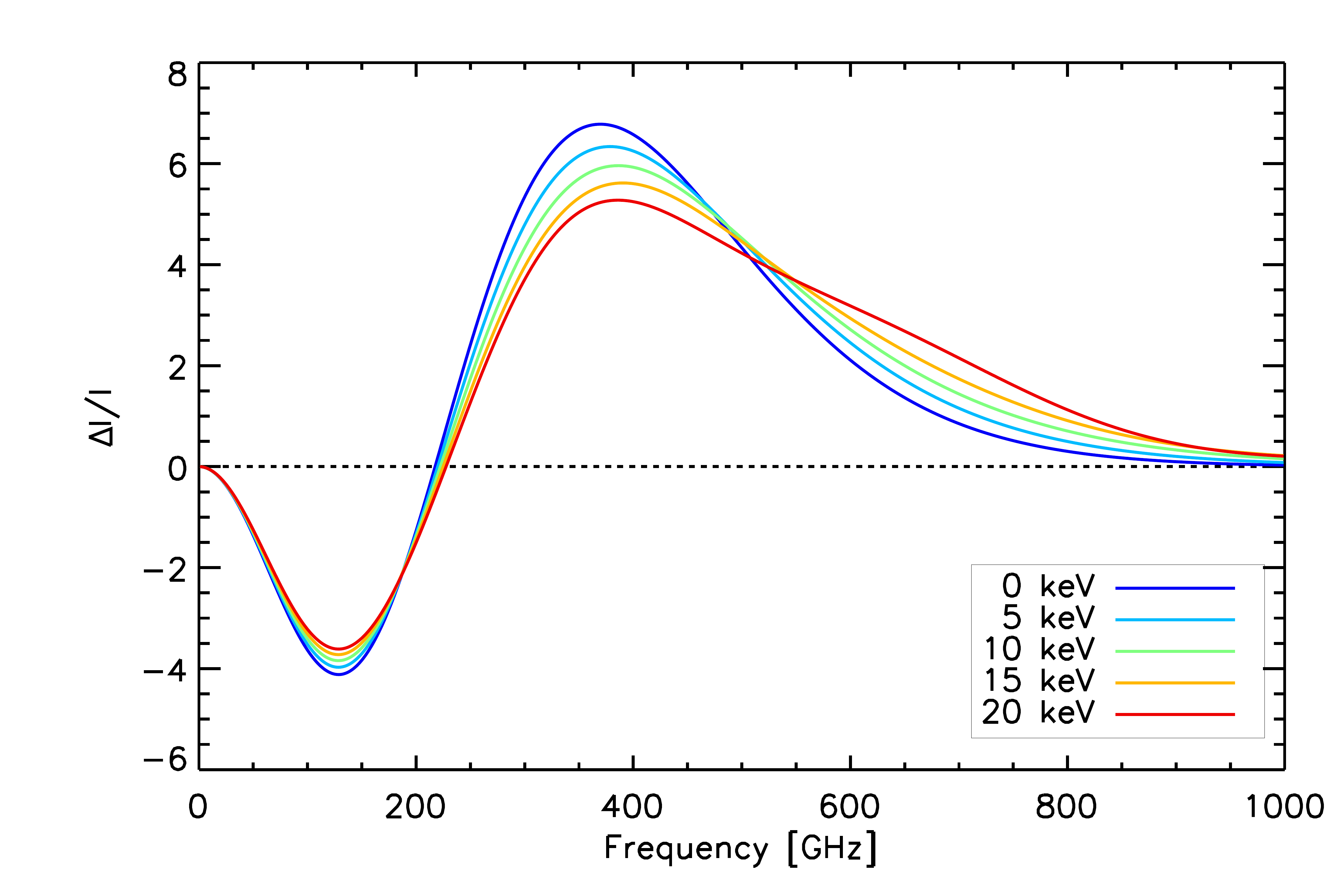}
\caption{tSZ spectral distorsion as a function of the frequency for various temperatures of the hot plasma from 0 to 20 keV.}
\label{szr}
\end{figure}

Figure~\ref{szr} shows the tSZ spectral dependance as a function of the frequency for various temperatures of the hot plasma ranging from 0 to 20 keV. We observe that the main consequence of relativistic corrections is a modification of the zero frequency, $\nu_0$, of the tSZ spectral distorsion, that follows the relation $\nu_0 \simeq 217.4 + T_e/2$. We also observe a significant increase of the 353 to 545 GHz tSZ intensity ratio. In general, higher temperatures for the plasma will favor a higher tSZ amplitude at high-frequencies, and a lower tSZ intensity at low frequencies. 

The Planck experiment has a large frequency coverage at low frequency ($<217$~GHz) where the tSZ effect produces an intensity decrement, at $217$~GHz where tSZ effect is almost null, and at higher frequencies ($>217$~GHz) where tSZ produce positive anisotropies on the CMB. This makes Planck HFI a really tailored instrument for the tSZ effect detection and scientific exploitation.

\section{Methodology}
\label{sec_meth}

\subsection{Estimation of tSZ flux per frequency and stacking}

We refer to \citet{hur14} for a detailed description of the tSZ flux extraction in Planck intensity maps. 
We first extracted 2x2$^{\rm o}$ patches around each galaxy clusters. We cleaned for infrared emission using the 857 GHz channel. We also computed a tSZ $y$-map with the MILCA method \citep{hur13} and we estimated the tSZ flux in each Planck frequency using the MILCA map as a template to build the spectral energy distribution (SED) toward each galaxy clusters.
Then, we subtract the 217 GHz signal to other channels in order to clean for CMB emission.
Finally, we divided our galaxy clusters samples into $T_e$ bins. We separated the MCXC into five temperature bins ($\Delta T_e = 2$ keV) and the spectroscopic sample into three temperature bins ($\Delta T_e = 4$ keV).
Then, we performed a stack of individual galaxy cluster maps for each temperature bin.

\begin{figure*}[!th]
\center
\includegraphics[width=18cm]{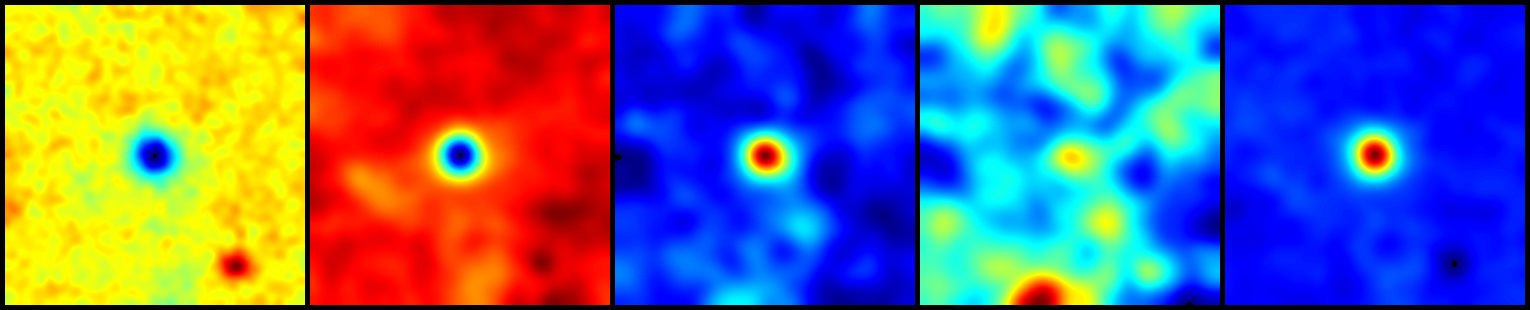}\\[-0.14cm]
\includegraphics[width=18cm]{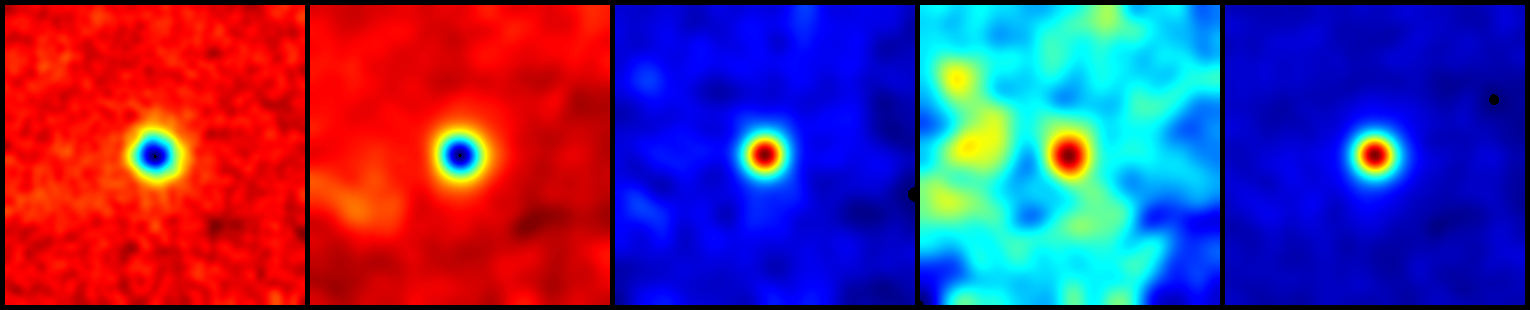}
\caption{From left to right: stack of Planck intensity maps at
  100, 143, 353, 545 GHz cleaned by the 857 and 217 GHz channel, and MILCA tSZ map, centered on the
  location of Planck MCXC clusters for a low-T$_e$ bin (top panel) and a high-T$_e$ bin (bottom panel). Each stacked map represents an area
  of 2$^\circ \times$ 2$^\circ$.}
\label{stack1}
\end{figure*}

Figure~\ref{stack1} shows the results of the stacking procedure toward MCXC clusters for the lowest temperature bin (top panel) and for the highest temperature bin (bottom panel).
We observe a significant amount of tSZ signal in the 545 GHz stacked map for the high-temperature bin, whereas the low-temperature bin does not present a significant tSZ emission at this frequency.

\subsection{Background and foreground contamination}
\label{errstat}
To estimate the contamination by other sources of anisotropies in Planck frequency maps, we performed the flux extraction, as described in \citet{hur14}, at 1000 random positions across the sky. These random positions follow the same spatial repartition in latitude as our sample of galaxy clusters to avoid any bias coming from the galactic plane area, which does not contain galaxy clusters in the used sample. In the following we consider that uncertainties are uncorrelated between two different galaxy clusters.\\
Consequently we can derived the full covariance matrix for the flux estimation in frequency channels from 100 to 545 GHz, an example of such a correlation matrix is presented in Table.~\ref{covmat}.
 This covariance matrix has a determinant of $2.4~10^{-4}$, which quantifies the volume occupied by the swarm of the data samples in our four dimension subspace (HFI frequencies from 100 to 545~GHz excluding 217~GHz).\\
Considering that most foreground and background emissions have been cleaned from the maps, the covariance matrix is dominated by the instrumental noise, that is correlated between frequencies due to the cleaning process. The high level of correlation between 353 and 545~GHz also receives contributions from thermal dust residuals in these maps.
We stress that this correlation matrix only accounts for uncertainties produced by uncorrelated components with respect to the tSZ effect.

\begin{table}
\caption{Correlation matrix of statistical uncertainties for emission tSZ emission law, estimate from 1000 random positions across the sky.}
\label{covmat}
\center
\begin{tabular}{|c|c|c|c|c|}
\hline
Frequency (GHz) & 100 & 143 & 353 & 545 \\ \hline
100 &1  &   0.62    &  -0.49     & -0.48 \\ \hline
143 &0.62    &   1     &   -0.49     & -0.56 \\ \hline
353 &  -0.49     & -0.49    &   1   &   0.83 \\ \hline
545 &    -0.48     & -0.56  &  0.83  &     1 \\ \hline
\end{tabular}
\end{table}

\subsection{Correlated foreground contamination}

As we discuss in Sect~\ref{errstat}, errors induced by uncorrelated foregrounds can be fairly estimated using random positions over the sky. However this estimation does not account for extra-noise and bias produced by physically correlated emissions, such as radio sources contamination at low frequency, cosmic infrared background (CIB) contamination at high frequency, and CMB secondary anisotropies themselves through the kSZ effect.\\

\subsubsection{Cosmic infrared background}
For the estimation of $T_{e}$ from tSZ measurement, the 545 GHz map is the key frequency channel, however it is also a frequency for which one the detection of the tSZ effect is the most challenging. 
At this frequency 10\% of the CIB emission is produced by objects with redshift lower than 1.0 \citep{lag05,add12}. See \citet{planckszcib}, for a measurement of the tSZ$\times$CIB cross correlation. An excess of emission at high frequency is produced by the CIB residuals, which thus mimic the effect of relativistic corrections on the tSZ spectral distorsion.

The tSZ emission scales as $M_{500}$ to the power of $1.79$ \citep{PlanckSZC} and the CIB scales as $M_{500}$ to the power of $\simeq 1.00$ \citep{planckszcib}. Consequently, the tSZ-CIB ratio decrease with $M_{500}$ and thus with $T_{500}$ as $F_{\rm CIB} \propto T^{-1.20}_{500}$ assuming that $T_{500}$ evolves as $M^{2/3}_{500}$. 
This implies that CIB residuals will be more important for low temperature galaxy clusters.
Consequently, CIB contamination can be separated from tSZ relativistic corrections by considering several bins of temperature. \\
We also stress that our dust-cleaning procedure using the 857 GHz channel will remove most of the CIB contamination, considering that the clusters in our sample are essentially low-$z$ objects \citep{planckszcib}. Indeed, dusty galaxy emission in these galaxy clusters presents a very similar spectral behavior to the Milky Way.

\subsubsection*{4.3.2. Radio point sources}

In order to avoid contamination by radio loud active galactic nucleus (AGN), we removed from the analysis clusters that present an emission above 0.5 mK$_{\rm CMB}$ at 100 GHz in a radius of 30' from the galaxy cluster position.

\subsubsection*{4.3.3. kSZ}
The kSZ effect follows the same spectral dependance as the CMB and thus is suppressed from our analysis by the CMB cleaning performed with the 217 GHz channel.
However, some kSZ residuals may remains in the measured SED due to calibration uncertainties \citep{planck2015cal}. The inter-calibration uncertainties at 100, 143, 217, 353, and 545~GHz are 0.09, 0.07, 0.16, 0.78, and 5\% respectively.
We stress that absolute calibration uncertainty does not affects the tSZ spectral signature, it only affects the overall Compton parameter normalization, which did not impact the estimation of $T_e$.\\
Propagating these uncertainties through our data processing, we deduced that the calibration uncertainties induce a leakage of the kSZ effect amplitude into tSZ effect flux measurement with a standard deviation of 0.2\%. The kSZ is typically one order of magnitude fainter than the tSZ effect, similarly tSZ relativistic corrections modify the tSZ spectral distorsion by $\simeq 10\%$. Consequently, kSZ effect residuals may affect the tSZ relativistic corrections measurement at $\simeq 0.2\%$ for a single galaxy cluster. Additionally, the kSZ effect is averaging to zero when stacking galaxy clusters, the bias for a given temperature bin is thus $\simeq 0.2\%/N_{\rm cl}$. Consequently, the kSZ component contamination can be safely neglected.

\subsection{Systematics produced by bandpass and calibration uncertainties}

The only Planck channel that presents a significant relative bandpass uncertainty for the tSZ effect is the 217 GHz channel \citep{PlanckBP}. However, as we clean for CMB using this channel, we are not sensitive to this uncertainty in the measurement of $T_e$ through relativistic corrections to the tSZ effect spectral distorsion.\\
However, the relative calibration uncertainty is a major limitation, as this uncertainty is higher at high frequency \citep{PlanckCAL}, where the tSZ relativistic corrections present a significant departure form non-relativistic tSZ.
Calibration uncertainties are  $<0.2$\% for 100, 143 and 217 GHz channels,  $<1\%$ at 353 GHz and 5\% at 545 and 857 GHz.
In the following, we consider uncertainties from calibration and propagate them to our analysis.

\section{Data analysis}
\label{sec_ana}
\subsection{Profile likelihood analysis}

To describe our measurement, the most general model reads
\begin{equation}
F^j_i = Y^j A_i(T_e) + F^j_{\rm sync}A_i^{\rm sync},
\label{specmod}
\end{equation}
with $A_i(T_e)$ being the exact tSZ spectral distorsion for a plasma temperature, $T_e$, $Y^j~=~\int y {\rm d}\Omega$ the integrated Compton parameter for temperature bin $j$, 
and $A_i^{\rm sync}$ a synchrotron spectrum with spectral index -1. Adjustable parameters are ${Y^j}$, $T_e$, and $F^j_{\rm sync}$ the synchrotron amplitude.\\

To fit the value of $T_e$ in each temperature bins, we used a profile likelihood approach. We use a flat prior for the synchrotron contamination: $0\%<F^j_{\rm sync}<15\%$. For each value of $T_e$ and $F^j_{\rm sync}$, we compute through an unbiased linear fit the tSZ flux, $\widehat{Y}^j$, of our measurement.\\
In this analysis we have uncertainties on both the measurement (mainly the CMB contamination) and the model (bandpass uncertainties), this two sources of uncertainties have similar amplitudes. Consequently we use the following estimator:
\begin{equation}
\widehat{Y}^j = \left[ {\bold A}^{T}{\cal W}{\bold A} - {\mathrm{Tr}({\cal C}_{A}^{T}{\cal W}}) \right]^{-1}\left[ {\bold A}^{T}{\cal W}\widehat{\bold F}^j \right],
\end{equation}
With ${\bold A}$ the tSZ transmission vector, ${\cal C}_{A}$ the ${\bold A}$ covariance matrix, $\widehat{\bold F}^j$ is the measured tSZ emission law and ${\cal W} = {\cal C}^{-1}_{F^j}$ is the inverse of the noise covariance matrix on $\widehat{\bold F}^j$.
Then we compute the $\chi^2$, for each couple of parameters ($T_{\rm CMB}$, $F^j_{\rm sync}$) as
\begin{equation}
\chi^2 = \left(\widehat{\bold F}^j - {\bold F}^j \right)^T \left({\cal C}_{F^j} + \widehat{Y}^2 {\cal C}_A \right)^{-1} \left(\widehat{\bold F}^j - {\bold F}^j \right).
\label{eqfit}
\end{equation}
Finally, we estimate the value of $T_e$ by marginalizing over $F^j_{\rm sync}$ and computing the first-order momentum of the likelihood function, ${\cal L} = e^{-\chi^2/2}$, with respect to $T_e$.  
We compute the uncertainties on $T_e$ using the second-order momentum of ${\cal L}$.

\subsection{Averaged electronic temperatures}

If we assume a first order approximation of the tSZ spectral distorsion, then the bin-averaged temperature is given by,
\begin{align}
T^{j}_e \simeq \frac{\sum_k T^{k}_e Y^{k}}{\sum_k Y^{k}},
\end{align}
where $T^{j}_e$ is the averaged temperature for the bin $j$, $T^{k}_e$ is the temperature of a given galaxy cluster, $k$, in the bin $j$, and $Y^{k}$ the integrated Compton parameter of the galaxy cluster $k$.
However, tSZ relativistic corrections are not linear with respect to $T_e$, thus $T^{j}_e$ differs from the Compton parameter weighted average of $T^{k}_e$. \\
We estimated the discrepancy between the Compton parameter weighted temperature and the real average by performing stacks of the real tSZ spectral distorsions in temperature bins, for electronic temperatures ranging from 0 to 15 keV and bin width ranging from 1 to 10 keV. Then, we fit for a single temperature in each temperature bin using Eq.~\ref{eqfit}. We stress that the bias is thus dependent on the experimental setting (frequency coverage and data covariance matrix).
We found that in small $T_e$ bins this discrepancy is negligible, below $0.05$ keV for bin width $\Delta T_e < 5$ keV. This supports the first order linear assumption when performing a stacking analysis of the tSZ relativistic corrections.

\subsection{Results}

\begin{figure}[!th]
\center
\includegraphics[width=9cm]{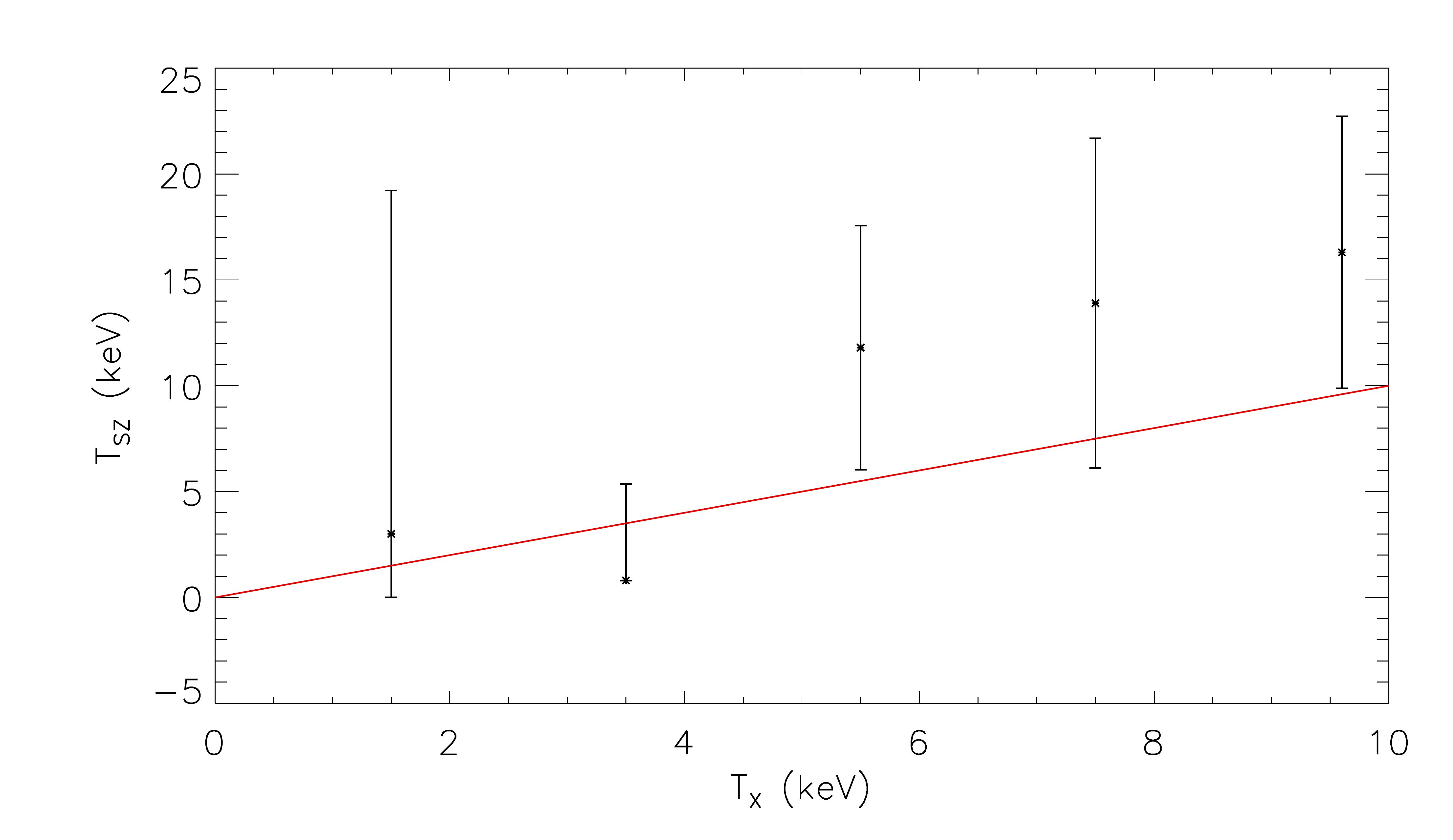}
\caption{Measured temperature from tSZ relativistic correction for the MCXC galaxy cluster sample, as a function of the temperature derived from X-ray luminosity. The 1:1 relation is shown as a solid red line.}
\label{stack0}
\end{figure}

Figure~\ref{stack0} shows the relation between the temperature derived from tSZ relativistic corrections and the temperature derived from X-ray luminosity. Fitting for the slope of this relation we obtain $T_{e,{\rm tSZ}} = (1.65 \pm 0.45) T_{e, {\rm X}}$, with a significance of 3.7 $\sigma$ and consistent with a 1:1 relation at 1.4 $\sigma$.
Considering the large number of galaxy clusters inside each bin, the uncertainties over the X-ray temperature average is small compared to temperature bin width and tSZ temperature uncertainties. Thus, we did not display these uncertainties.

\begin{figure}[!th]
\center
\includegraphics[width=9cm]{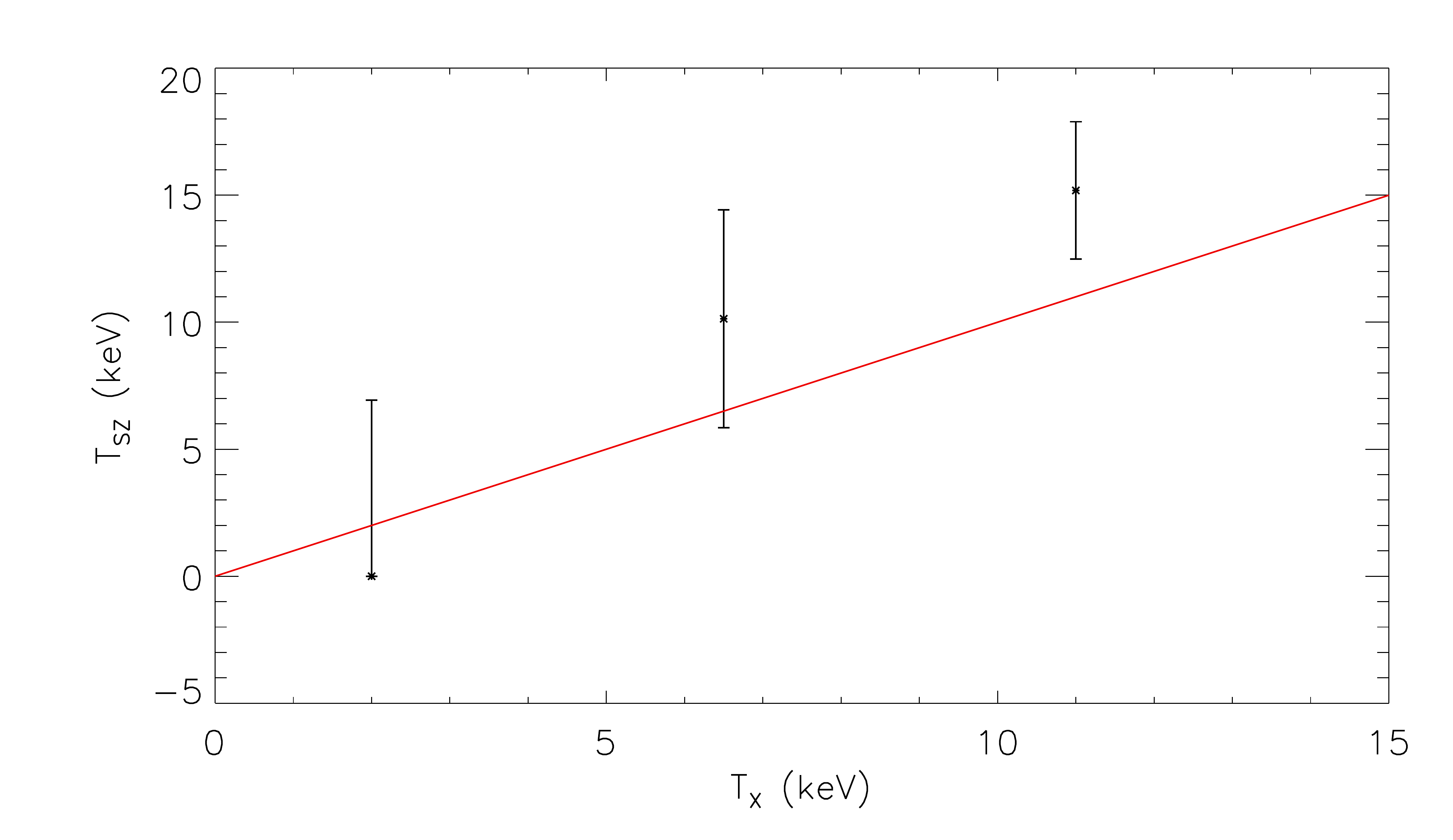}
\caption{Measured temperature from tSZ relativistic correction for the galaxy clusters with X-ray spectroscopic temperature, as a function of the temperature derived from X-rays. The 1:1 relation is shown as a solid red line.}
\label{stack2}
\end{figure}

Figure~\ref{stack2} shows the relation between the temperature derived from tSZ relativistic corrections and the temperature derived from spectroscopic X-ray analyses. Fitting for the slope of this relation we obtain, $T_{e,{\rm tSZ}} = (1.38 \pm 0.26) T_{e, {\rm X}}$, with a significance of 5.3 $\sigma$ and consistent with a 1:1 relation at 1.5 $\sigma$.
Uncertainties over the X-ray spectroscopic temperature are small compared to the bin width and tSZ temperature uncertainties. Consequently, they are displayed on the figure.

In both cases, Figs.~\ref{stack0}~and~\ref{stack2}, we recover the expected relation between tSZ estimated temperatures and X-ray temperatures. We do not observe significant contamination produced by CIB residuals, that would appears as an overestimation of the tSZ temperature in the lowest temperature bins. We observe in both cases that the tSZ temperature for the highest X-ray temperature bins overestimates the X-ray temperature. This bias might be produced by calibration uncertainties leading to an overall bias in all temperature bins. In Fig.~\ref{stack2}, the highest temperature bin is dominated by calibration uncertainties, which supports the calibration origin of the observed high-temperature overall excess for the tSZ relativistic corrections.

\section{Discussion and conclusion}

We have performed the first high signal-to-noise ratio detection of the tSZ relativistic corrections at a significance level of 5.3 $\sigma$.
We have considered potential contamination by radio and infrared emission and shown that thes sources of contaminations are negligible.\\
In this work, the tSZ relativistic corrections detection has been achieved through a statistical analysis of large galaxy cluster samples. In particular, we used several temperature bins to distinguish real tSZ relativistic corrections signature from systematic effects such as CIB contamination or calibration uncertainties.
This analysis exhibits the complexity of the tSZ relativistic corrections recovery from a small number of frequencies due to the required cleaning of CMB and infrared astrophysical components.\\
However, future CMB-experiments, such as COrE+\footnote{\url{http://hdl.handle.net/11299/169642}}, which will have better sensitivity or a more refined frequency coverage will offer the possibility of performing better measurements, and scientific exploitation of the tSZ relativistic corrections ({\color{blue} Hurier et al., in prep}).\\
The tSZ effect spectral distorsion is commonly used as a prior for galaxy cluster detection, this analysis shows that we are now reaching the level of accuracy where tSZ relativistic corrections can be detected. This implies that galaxy cluster detection based on the tSZ spectral distorsion neglecting relativistic corrections might be biased toward low-temperature objects, and then bias the detection of high-z high-mass galaxy clusters. \\

\section*{Acknowledgements}
\thanks{The author thanks N.Aghanim \& D.Poletti for useful discussions and comments. We acknowledge the support of the French \emph{Agence Nationale de la Recherche} under grant ANR-11-BD56-015.

\bibliographystyle{aa}
\bibliography{relat_tsz_publi.bib}

\end{document}